\documentclass[prb,twocolumn,floatfix]{revtex4}
\usepackage{epsfig}
\usepackage{dcolumn}

\usepackage[latin1]{inputenc}
\usepackage[T1]{fontenc}

\begin{document}
\newcommand{\rar}{$\rightarrow$}
\newcommand{\lrar}{$\leftrightarrow$}

\newcommand{\beq}{\begin{equation}}
\newcommand{\eeq}{\end{equation}}
\newcommand{\bea}{\begin{eqnarray}}
\newcommand{\eea}{\end{eqnarray}}
\newcommand{\req}[1]{Eq. (\ref{Eq#1})}
\newcommand{\degree}{$^{\rm\circ} $}
\newcommand{\pcite}{\protect\cite}
\newcommand{\pref}{\protect\ref}
\newcommand{\rfg}[1]{Fig. \ref{F#1}}
\newcommand{\rtb}[1]{Table \ref{T#1}}

\title{Sequence-dependent B-A transitions in DNA {\em in silico}:
Electrostatic condensation mechanism}

\author{Alexey K. \surname{Mazur}}
\affiliation{CNRS UPR9080, Institut de Biologie Physico-Chimique,
13, rue Pierre et Marie Curie, Paris,75005, France\\
}
 

\begin{abstract}
Dynamics of the polymorphic A<->B transitions in DNA is compared for
two polypurine sequences, poly(dA).poly(dT) and poly(dG).poly(dC),
long known to exhibit contrasting properties in experiments. In free
molecular dynamics simulations reversible transitions are induced by
changing the size of a water drop around DNA neutralized by Na$^+$
ions. In poly(dG).poly(dC) the B<->A transitions are easy, smooth and
perfectly reversible. In contrast, a B->A transition in
poly(dA).poly(dT) dodecamer fragment could not be obtained even though
its A-form is stable under low hydration. Normal B->A transitions are
observed, however, in long poly(dA).poly(dT) stretches flanked by GC
pairs. An intermediate range of hydration numbers is identified where
opposite transitions are observed in the two dodecamer fragments,
namely, A->B in poly(dA).poly(dT) and B->A in poly(dG).poly(dC). With
hydration numbers close to the stability limit of the B-form, the two
sequences exhibit qualitatively different counterion distributions,
with a characteristic accumulation of Na$^+$ ions next to the opening
of the minor groove in poly(dA).poly(dT). This difference can explain
an increased persistence of poly(dA).poly(dT) DNA towards A-form in
crystalline and amorphous fibers as compared to solution conditions.
The good overall agreement with experimental data corroborates the
general role of the electrostatic condensation mechanism in the A/B
polymorphism in DNA.
\end{abstract}

\maketitle

\section*{Introduction}

Reversible transitions between the A and B forms of DNA
\cite{Saenger:84} represent one of the modes for governing protein-DNA
interactions, with the B-form as the dominant biological conformation
and the A-form as a high energy state adopted temporarily for
different purposes \cite{Ivanov:95,Lu:00}. These transitions can be
induced in several ways {\em in vitro}
\cite{Franklin:53,Leslie:83,Piskur:95,Tunis-Schneider:70} and they
were extensively studied in the recent decades. However, a number of
interesting issues are still unclear including the underlying
molecular mechanisms and the driving forces. Different theoretical
approaches were applied in this field during its long history
\cite{Ivanov:74,Calladine:84,Mazur:89,Hunter:93,Ivanov:79}.
In the recent years realistic molecular dynamics (MD) simulations of
A\lrar B transitions became possible including relevant explicit
solvent environment
\cite{Cheatham:96,Yang:96,Cheatham:97b,Cheatham:97c,Cieplak:97,Jayaram:98,Sprous:98,Mzjacs:03}.
Despite the limited duration of trajectories and some force field
artifacts \cite{Feig:98}, these studies helped to get new insights in
the putative molecular mechanisms of the DNA A/B polymorphism. A new
approach recently found in this field consists in using small water
drops as solvent shells around DNA fragments \cite{Mzjacs:02}. By
changing the size of the drop, reversible A\lrar B transitions can be
induced, with hydration numbers in rough agreement with experimental
values \cite{Mzjacs:03}.  

The water drop simulation approach is the first and the only presently
known technique to obtain reversible B\lrar A transitions in DNA {\em
in silico}. The established mechanism of such transitions essentially
consist in an intra-duplex electrostatic condensation of the two DNA
strands. With the drop size gradually reduced, the equilibrium is
shifted towards the A-form due to inversion of electrostatic
interactions across the major DNA groove when the local cation
concentration exceeds a certain transition level. The important role
of this particular type of interactions is well established in the
earlier literature \cite{Feuerstein:86,Rouzina:98}. Similar
B\rar A transitions were observed experimentally in poly(dG).poly(dC)
fragments with multivalent cations bound in the major groove
\cite{Xu:93a}, which could be reproduced in simulations
\cite{Cheatham:97c}. This mechanism is perhaps generally responsible
for the A/B polymorphism, but a number of other ideas concerning its
nature are also being discussed. Among them one can mention the factor
of economy of hydration of the sugar phosphate backbone
\cite{Saenger:86}, the base stacking interactions \cite{Calladine:84}
and the hydrophobic effect \cite{Ivanov:73,Hunter:93}. It is
interesting, therefore, to know if the A\lrar B transitions in water
drop simulations that are relatively simple to study and rationalize
are really similar to those in experimental conditions. To this end we
should check if different aspects of the A/B polymorphism {\em in
silico} correspond to the experimental results for the most part
obtained for DNA in fiber crystals under low relative humidity and in
concentrated solutions of some non-polar solvents.  Among such aspects
the sequence dependent nature of the A/B polymorphism is perhaps most
intriguing and well documented.

Sequence effects upon the B\rar A transition attract long-standing
interest. In fiber crystals, DNAs with different regular sequences
often exhibit dissimilar capacities to adopt the two forms
\cite{Leslie:83,Zimmerman:83b}. The same trends are observed in
solution, moreover, here the pattern revealed is much richer because a
large variety of sequences could be probed, with their propensities
towards A- and B-forms measured quantitatively \cite{Ivanov:95}. It
was shown that, during transitions in long natural DNA, different
regions switch between the two forms in a complex sequence specific
order \cite{Becker:89}. The A/B-philicity of different base pair steps
and triplets was examined systematically in synthetic fragments.  It
could be parametrized and used successfully for predicting the
properties of mixed sequences
\cite{Ivanov:95,Basham:95,Tolstorukov:01}. All these effects certainly
originate from interactions between stacked bases and some of them
could be rationalized \cite{Calladine:84,Mazur:89,Hunter:93}.

The measured A/B propensities appear such that consecutive base pair
steps tend to compensate one another, therefore, generic DNA sequences
commonly are neither A- nor B-philic \cite{Tolstorukov:01}.  A strong
difference is observed, however, between the stretches of consecutive
guanines and adenines (G-tracts and A-tracts, respectively). Double
helical DNA fragments with these two sequences exhibit contrasting
physico-chemical properties as established by a number of different
methods
\cite{Peck:81,Rhodes:81,Strauss:81,Hogan:83,Leslie:83,Zimmerman:83b,Nishimura:86,Behling:86,Benevides:86,Sarma:86,Peticolas:88,Buckin:89a,Buckin:89b,Brahms:92,Chalikian:94,Vorlickova:96,Ivanov:96}.
A systematic review of these data has been published
recently\cite{Hud:03}. In G-tracts, the B\rar A transition is very
easy. The poly(dG).poly(dC) DNA is generally prone to adopt the A-DNA
conformation in conditions where random sequences stay firmly in the
B-form \cite{Leslie:83,Nishimura:86,Peticolas:88}. In contrast, in
A-tracts, the B\rar A transition is particularly difficult
\cite{Leslie:83,Peticolas:88,Ivanov:96}. Notably, a B\rar A transition
was never observed in fiber crystalline poly(dA).poly(dT) because in
standard transition conditions it is reportedly prone to denaturation
or alternative transformations \cite{Leslie:83,Zimmerman:83b}. Solvent
cations were found to bind in the major grooves of G-tracts and in the
minor grooves of A-tracts and not vice versa, which is probably
related and perhaps responsible for the A/B propensities of these
sequences \cite{Hud:01,Hud:03}. This remarkable physico-chemical
difference is particularly important in the context of the intriguing
roles of A-tracts and G-tracts in the genome functioning.  Various
A-tract sequences are overrepresented in both procariotic and
eucariotic genomes \cite{Dechering:98}. They are long known to cause
DNA curvature in solution that can affect binding to proteins and
formation of nucleosomes \cite{Trifonov:80,Widlund:97}.  Long G-tracts
are also widespread in eucariotic as well as archaebacteria genomes
\cite{Vashakidze:87}.

\begin{figure*}
\centerline{\psfig{figure=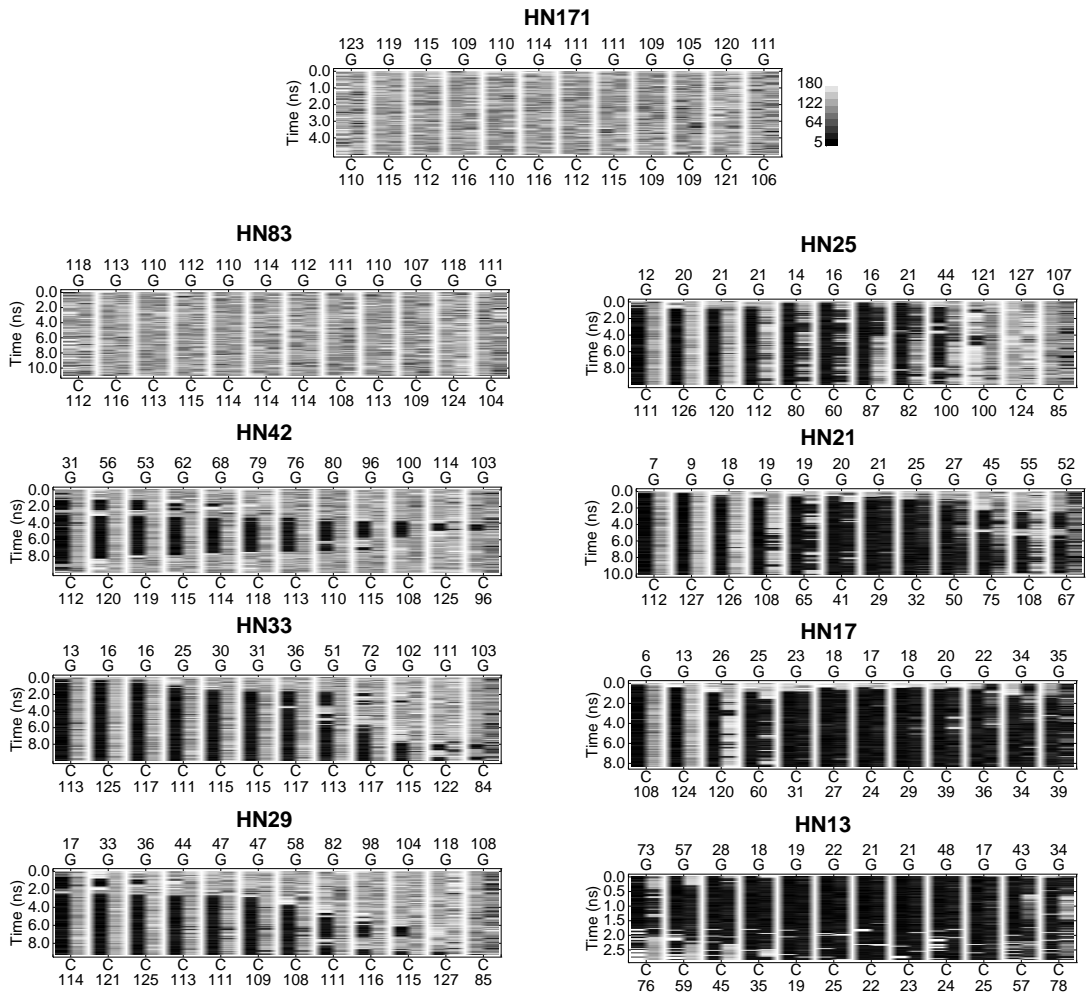,width=15.0cm,angle=0.}}
\caption{\label{Fg12sgph}
Dynamics of sugar pucker pseudorotation \cite{Altona:72} in G12 under
different hydration numbers.  The boundary pucker values (ca 0\degree and
180\degree) are assigned the black and the white colors, respectively,
with intermediate values mapped linearly to the gray scale levels.
Each base pair step is characterized by a column consisting of two
sub-columns, with the left sub-columns referring to the sequence
written above in 5'-3' direction from left to right with the time
averaged phases given on top. The right sub-columns refer to the
complementary sequence shown below together with the corresponding
time averaged phases. All trajectories shown in this figure started
from the B-DNA state except the HN13. The last simulation started from
the final state of HN17 because the original B-DNA state is unstable
and prone to collapse under that low hydration.}\end{figure*}

Here we report about the results of systematic simulation studies of
A\lrar B transitions in a series of DNA fragments with these two
characteristic sequence motifs. The established experimental
preferences of A- and G-tracts towards B and A forms, respectively,
are qualitatively well reproduced. In poly(dG).poly(dC) dodecamer, the
B\rar A transition is easy, smooth and perfectly reversible. In
contrast, a B\rar A transition in similar poly(dA).poly(dT) fragment
could not be obtained even though its A-form was found stable under
low hydration. At the same time, normal B\rar A transitions are
observed in long poly(dA).poly(dT) stretches flanked by GC pairs. An
intermediate range of hydration numbers is found where opposite
transitions are observed in the two dodecamer fragments, namely, A\rar
B in poly(dA).poly(dT) and B\rar A in poly(dG).poly(dC).  These
results agree with the assumed similarity of the A\lrar B transitions
in experiment and in water drop simulations. It appears that, with
hydration numbers close to stability of the B-form, these two
fragments have qualitatively different counterion distributions, with
a characteristic accumulation of Na$^+$ ions next to the opening of
the minor groove in poly(dA).poly(dT). In fibers, these ion positions
can be additionally stabilized by intermolecular contacts, which would
explain an increased persistence of poly(dA).poly(dT) DNA towards
A-form in crystalline and amorphous fibers as compared to solution
conditions.

\section*{Results}

\subsubsection*{Comparative dynamics of A\lrar B transitions}

\begin{figure}
\centerline{\psfig{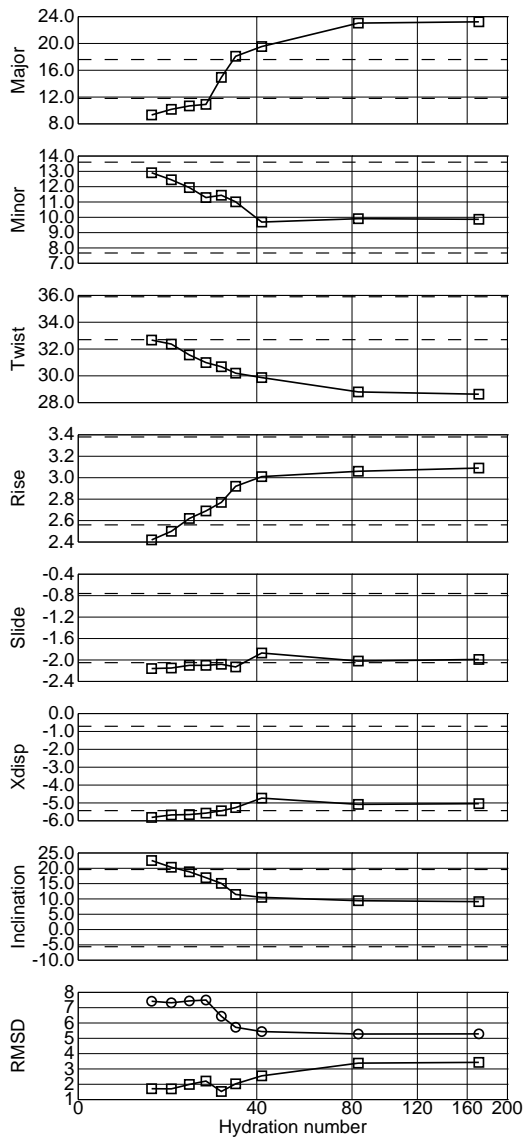}}
\caption{\label{Fg12alpa}
Quasi-static pattern of A\lrar B transitions for G12 as monitored by
different structural parameters. For each hydration number, a long MD
trajectory was computed and the last 1 ns average structures were
analyzed to obtain the necessary values. The two top plates show the
average groove widths measured as described elsewhere
\cite{Mzjmb:99,Mzjacs:03}. Several helical parameters are all computed
by Curves \cite{Curves:}, with the local values for the Slide and
global ones for the rest. The RMSD from the canonical A and B-DNA
conformations is shown by circles and squares, respectively.  All
distances are in angstr{\"o}ms and angles in degrees. The dotted lines
indicate canonical A- and B-DNA levels.}\end{figure}

\rfg{g12sgph} shows the sugar pucker dynamics in a series of
representative trajectories of G12.  Transitions from C2'-endo to
C3'-endo are observed starting already from HN42. At first they occur
only in the purine strand and for HN33 and HN29 an intermediate state
is obtained with this strand almost entirely in the C3'-endo
conformation. A complete transition to A-form is reached in the middle
of the duplex starting from HN25. This transition midway point is
similar to that reported for the dodecamer CGCGAATTCGCG
\cite{Mzjacs:03}. However, here the A-DNA backbone appears at much
higher hydration numbers and, as a result, the transition looks
smoother and less cooperative. Moreover, the final A-DNA structure
spreads from the center almost to entire fragment so that the
B-philicity of the DNA ends, generally observed in experiments and
calculations \cite{Minchenkova:86,Mzjacs:03}, is less significant. All
this agrees well with the known A-philicity of the poly(dG).poly(dC)
DNA \cite{Note1}.

\rfg{g12alpa} shows a titration-like pattern of these transitions as
followed by different structural parameters. Different traces all
exhibit S- or Z-shaped profiles sometimes with a very distinct
transition zone \cite{Note2}. The A-DNA structures observed under low
hydration are very close to the canonical conformation, with final
RMSD values below 2 \AA. In contrast, under the highest hydration, an
underwound B-DNA is observed with a strong bias towards A-form.
Similar deviations were reported earlier by others
\cite{Trantirek:00,Stefl:01,Lankas:02} and they are partially due to
the known negative forcefield bias of the Twist \cite{Cheatham:99}. The
A\lrar B transitions for G12 are easily achieved in both directions
and the points in \rfg{g12alpa} corresponding to the transition zone
could be reproduced in alternative simulations starting from the A-DNA
state.

\begin{figure*}
\centerline{\psfig{figure=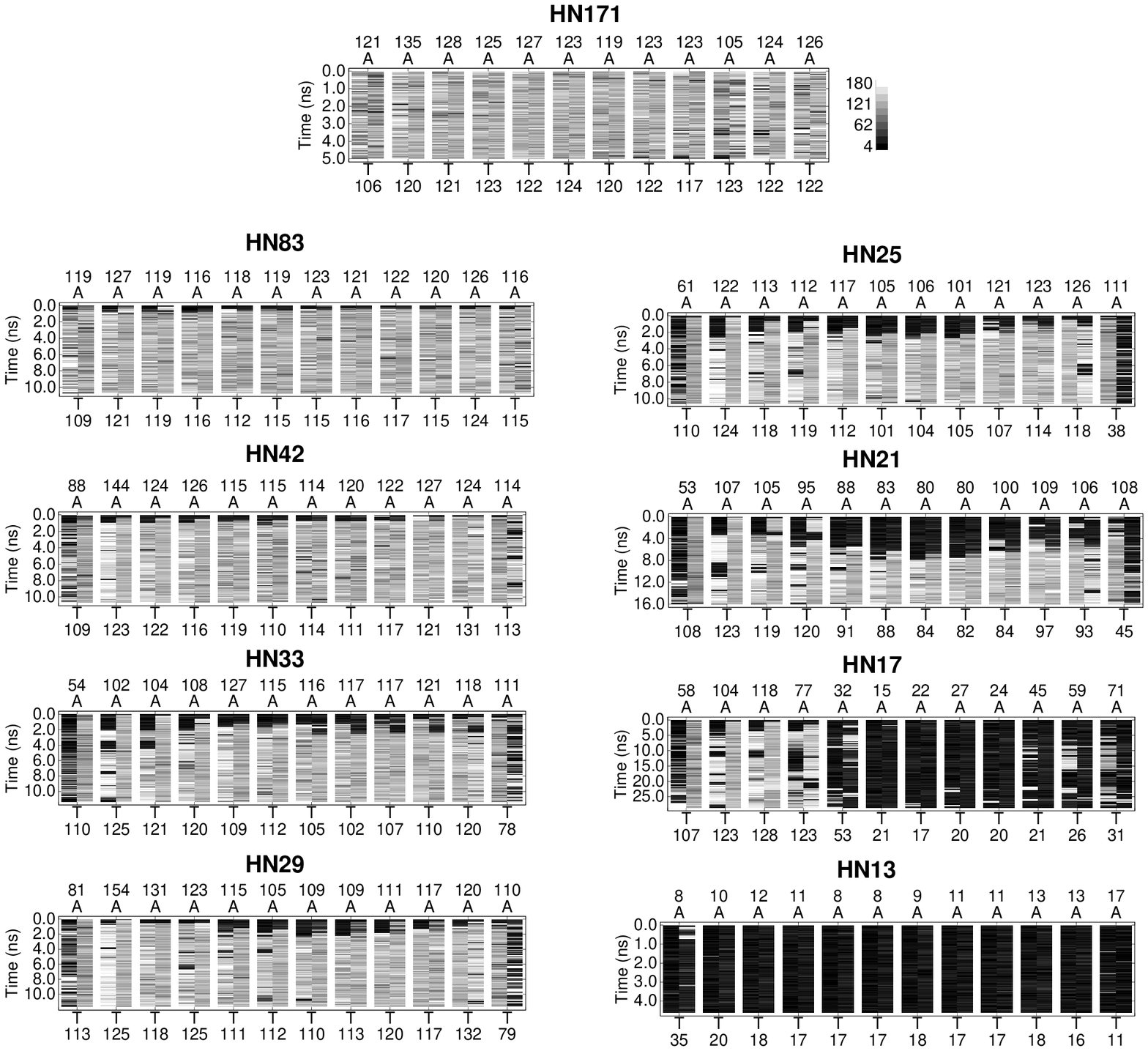,width=16.8cm,angle=0.}}
\caption{\label{Fa12sgph}
Dynamics of sugar pucker pseudorotation in A12 under different
hydration numbers. All trajectories shown in this figure started from
the A-DNA state .}\end{figure*}

\begin{figure}
\centerline{\psfig{figure=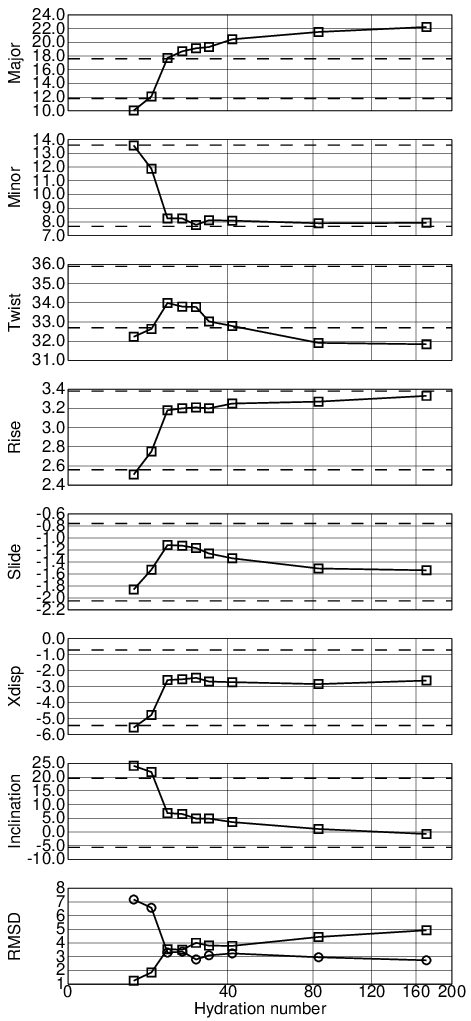,width=8.4cm,angle=0.}}
\caption{\label{Fa12alpa}
Quasi-static pattern of A\lrar B transitions for A12. The notation
is as in \rfg{g12alpa}.
}\end{figure}

Similar results for A12 are shown in the next two figures. The most
striking difference from G12 could not be shown here, namely, that we
failed to obtain a B\rar A transition for this fragment in spite all
efforts. At the same time, the A-form was perfectly stable under low
hydration and, for HN21 and higher numbers, the A\rar B transitions
passed without problems. Therefore, for \rfg{a12sgph}, the
trajectories chosen all started from the A-DNA conformation and we may
add that, with the exception of HN13 and HN17, the B-DNA dynamics was
also stable and converged to the same structures. Comparison of
\rfg{a12sgph} with \rfg{g12sgph} readily reveals a stretch of hydration
numbers were opposite transitions are observed in G12 and A12, that is
the simulations reproduce the expected relative difference between
these two fragments as regards their preference towards A- vs B-DNA.

\rfg{a12alpa} exhibits a pattern qualitatively similar to that in
\rfg{g12alpa} with a few exceptions. Again the computed A-DNA
structures are very close to the canonical conformation, but in this
case the computed B-DNA are also relatively similar to the canonical
one, even though the twist remains strongly underestimated. These
traces were obtained by increasing hydration as shown in
\rfg{a12sgph}. In the opposite sense they could be also reproduced
except for the hydration numbers below 21. In conditions HN17 and
HN13, repetitive trajectories of A12 starting from B-DNA resulted in
opening of terminal base pairs followed by irregular deformations that
have been interpreted as beginning of denaturation. Interestingly,
when this limiting hydration is approached from the right in
\rfg{a12alpa} the data show no trend towards the A-form. Instead some
parameters are nearly stable while the other even exhibit a trend in
an opposite direction, suggesting that the B-DNA structure is
progressively stabilized before the eventual base pair opening and
collapse.

\subsubsection*{Evolution of Na$^+$ distributions around DNA}

\begin{figure}
\centerline{\psfig{figure=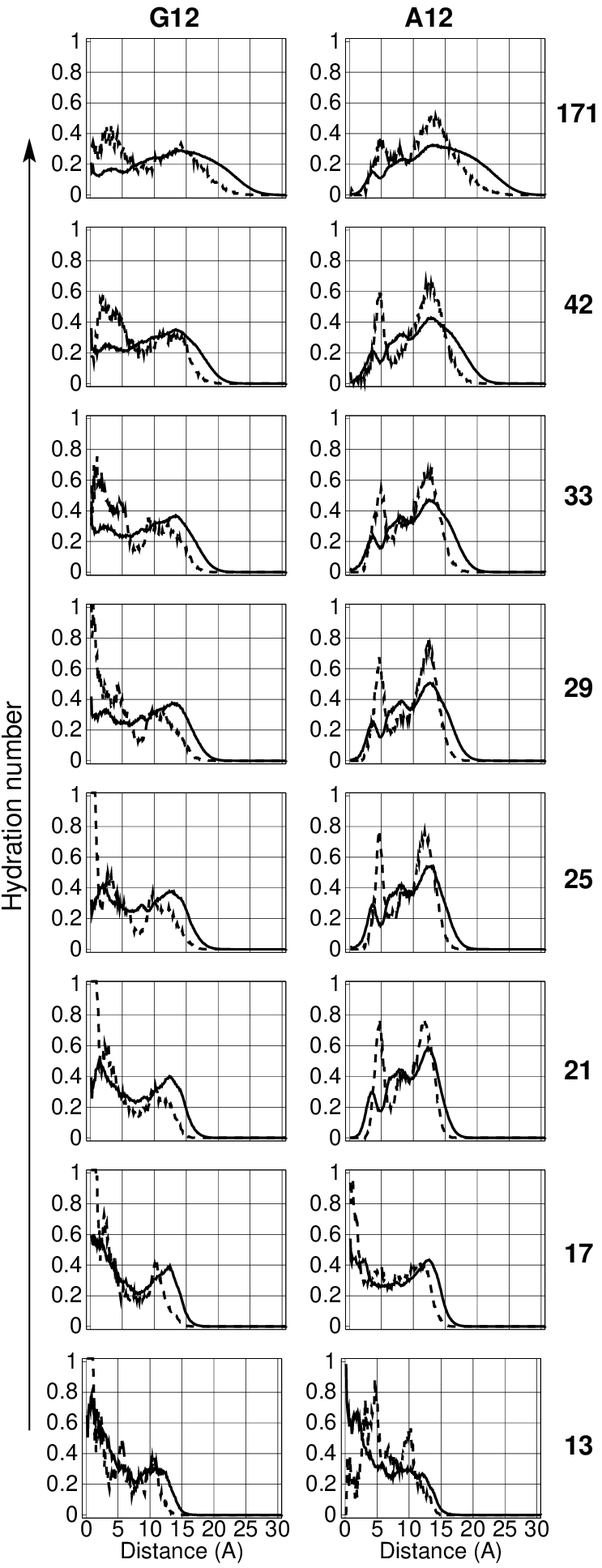,width=6.5cm,angle=0.}}
\caption{\label{Falhs}
Cylindrical radial distribution functions for water oxygens (solid
lines) and Na$^+$ ions (dashed lines) around G12 and A12.  DNA
structures saved in the last nanosecond of dynamics together with
surrounding water and counterions were superimposed with the canonical
canonical B-DNA with the global coordinate OZ direction as the common
helical axis. The Na$^+$ ions and water oxygens were counted in
coaxial 0.1 \AA\ thick cylinders.  The distributions are volume
normalized, that is scaled with a factor of $1/r$, and the final plots
were area normalized. The corresponding hydration numbers are shown on
the right. }\end{figure}

The distributions of counterions around DNA are compared for G12 and
A12 in \rfg{alhs}. The patterns shown here are interpreted as follows.
The duplexes are entirely covered by water, therefore, the radial
distribution of water oxygens shows the evolution of the available
volume not occupied by DNA. Even though the relative sizes of water and
Na$^+$ are not identical, these profiles give a rough estimate of free
space that can be sampled by ions, and, in the absence of specific
DNA-ion interactions, the solid and dotted traces in \rfg{alhs} should
have similar peak positions. In contrast, radial zones of strong
Na$^+$-DNA interaction produce strong separate peaks.  Note also that
the height of each peak in a cylindrical distribution should be
multiplied by the corresponding distance when their relative weights
are estimated.

The characteristic distributions for B-DNA are best seen in the
several top plates of A12. It does not differ from similar
distributions earlier reported for CGCGAATTCGCG \cite{Mzjacs:03}. The
three broad water peaks at approximately 4, 8 and 12.5 \AA\
correspond, respectively, to the first water layer in the major
groove, the next few layers in both major and minor grooves and the
bulk water outside DNA radius which is about 10 \AA\ for both A- and
B-DNA. The first Na$^+$ peak at 5 \AA\ is a sum of several different
contributions. It involves ions that interact with bases directly in
depth of the minor groove as well as those in contact with N7 atoms of
purine bases in the major groove. Mobile ions in the second hydration
shell in the major groove also contribute to this peak because Na-OW
distances are somewhat shorter than OW-OW ones. In cylindrical
distributions in \rfg{alhs} these qualitatively different
contributions could not be resolved and that is why Na$^+$ peaks at 5
\AA\ are found in B- as well as A-DNA distributions and often vary in
height. The outer peak at 12.5 \AA\ includes highly mobile ions around
DNA involved in non-specific phosphate screening. Its prominent part
that becomes less mobile under low hydration corresponds to Na$^+$
positions along the minor groove between the two phosphate strands.
Finally, an intermediate peak at approximately 8 \AA\ is generally
produced by mobile ions in the minor groove. Ion entering into the
minor groove was relatively rare, perhaps because the starting Na$^+$
distributions sampled only outer positions, which explains why such
peaks are less well reproduced between the trajectories.

The characteristic A-DNA distributions can be seen in the bottom
plates of A12 as well as in many plates of G12. In this case the
center of the DNA cylinder is filled with water and, consequently,
accessible for ions.  Most of them rest inside the DNA radius. As
regards the relative weights, the most prominent peak is found at 10
\AA\ and it corresponds to Na$^+$ sandwiched between the opposite
phosphate groups in the opening of the major groove. It is the
accumulation of counterions at this region that causes the B\rar A
transition \cite{Mzjacs:03}.

For G12 even under high hydration the Na$^+$ distribution is rather
different from typical B-DNA. These structures have a strong negative
Xdisp, therefore, the helical axis is shifted to the major
groove and is accessible to water and ions. Even though in HN82 most
of Na$^+$ stay outside the DNA radius their population inside the
major groove is already significant. With hydration numbers reduced,
the counterions are progressively pushed inside DNA. The Na$^+$ peak
corresponding to the opening of the major groove emerges gradually and
starting from HN25 it becomes predominant. All this corresponds well
to the relatively smooth B\rar A transition characterized above in
\rfg{g12sgph} and \rfg{g12alpa}. The A12 patterns in the right column
are radically different. In this case both water and Na$^+$
distributions retain the characteristic B-DNA shapes even with the
hydration number reduced to 21. The ions are not pushed inside DNA
when the outer water shell is reduced, and the Na$^+$ peak at 12.5
\AA\ remains dominant.

An evident demonstration of the distinction between these two DNA
fragments as regards their relative capacities to adopt A- and B-forms
is given in \rfg{snaps} where the snapshots are shown for A12 and G12
at the end of dynamics simulated in drops of the same size (HN25).
These structures and the counterion distributions confirm the main
aspects of dynamics illustrated by previous figures. For A12 we have a
B-DNA with a narrow minor groove near both ends and a widening in the
middle. A large number of the Na$^+$ ions stays outside DNA in front
of the minor groove. In contrast, G12 gives a typical A-DNA
conformation with a layer of Na$^+$ sandwiched between the opposite
phosphate groups in the narrow major groove.  This "electrostatic
sandwich" provides the main driving force of the B\rar A transition
according to the condensation mechanism
\cite{Cheatham:97b,Rouzina:98,Mzjacs:03}.

\begin{figure*}
\centerline{\psfig{figure=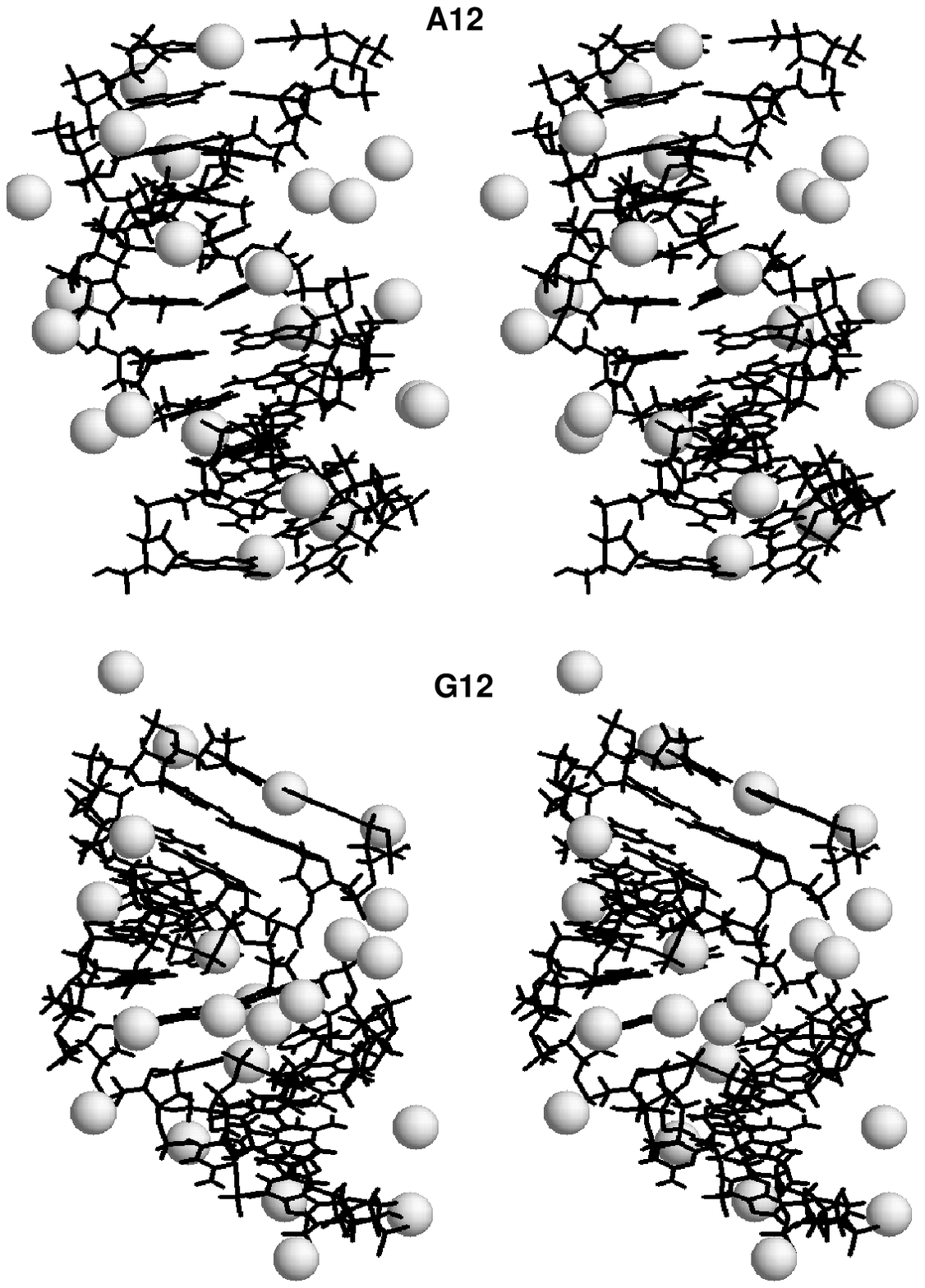,width=11.5cm,angle=0.}}
\caption{\label{Fsnaps}
Stereo snapshots of the final computed states of A12 and G12 obtained
with the same drop size corresponding to the hydration number of 25.
The Na$^+$ ions are shown by spheres.}\end{figure*}

\subsubsection*{Persistence of poly(dA).poly(dT) DNA to A-form}

A series of additional calculations was carried out in order to check
the origin of the A12 persistence with respect to A-DNA. It was
possible that, the  poly(dA).poly(dT) was intrinsically
unstable with respect to base pair opening during the B\lrar A
transition due to mechanical stretching of base pairs under
intermediate hydration numbers. In solution experiments, DNA commonly
exhibits reduced thermostability under intermediate alcohol
concentrations \cite{Ivanov:85}. It can even denaturate and then
renaturate when the water activity is further reduced
\cite{Vorlickova:91b}. To probe this transitional instability several
long trajectories were run for A-DNA close to the intermediate
hydration. Other simulations involved forced opening of one base pair
to check if this could induce denaturation. In all these tests the
A-DNA remained stable.

The opening of terminal base pairs could also result from kinetic
instability of B-DNA when it is suddenly transfered under low
hydration. Such an opening is sometimes observed even for G12 when
trajectories start from B-DNA in HN13. To check this possibility,
simulations were continued from the final B-DNA state of the HN21
trajectory, with the amount of water reduced smoothly in several steps
by removing each time only molecules not in contact with ions.  This
slow "evaporation" also led to opening of a terminal base pair under
an intermediate hydration, with the structure remaining in the B-form.

Finally, it was possible that the opening of terminal base pairs masks
the B\rar A transition kinetically, that is the molecule just cannot
stay intact long enough for the transition to occur. To check this the
two opposite terminal AT base pairs were replaced by GC and dynamics
was simulated starting from B-DNA in HN17 and HN13 water drops. In
these simulations no base pair opening was observed.  For HN17 two
trajectories were computed of 11 and 7 ns, respectively, with
different random initial distributions of counterions. In both cases
the duplex remained firmly in the B-DNA state. For HN13 the trajectory
was continued to 5.5 ns. During this time water started to leave the
minor groove and the duplex suffered irregular deformations that drove
it to a shrunk collapsed structure with partially closed grooves, and
yet the transition to the A-form did not occur.

The foregoing pattern is in a remarkable correspondence with the well
documented poly(dA).poly(dT) phobicity towards the A-form
\cite{Leslie:83,Zimmerman:83b}. Unfortunately, this finding does not
yet mean that a simple mechanism of this resistance can be readily
extracted from simulations. It appeared that the B\rar A transitions
in poly(dA).poly(dT) fragments become possible when the chain length
is increased. In a 16-mer fragment of poly(dA).poly(dT) under HN17,
opening of several terminal base pairs at both ends was also observed,
however, this was accompanied by a B\rar A transition in a few central
residues. When the two terminal base pairs at the opposite ends were
replaced with GC the resulting 16-mer fragment CGAAAAAAAAAAAAGC under
the same hydration showed a normal B\rar A transition in the middle
12-mer A-tract with the same general features as described before.
With the length of the A-tract further increased to 18 base pairs, a
B\rar A transition could also be obtained under HN21.

\section*{Discussion}

The experimental studies of the B/A polymorphism in solution earlier
revealed a surprisingly complex pattern of the double helical DNA
behavior depending upon its sequence as well as environment conditions
such as the water activity, the temperature, the type and the
concentration of counterions.  Perhaps not all effects observed {\em
in vitro} are biologically relevant, but the exceptional role of the
DNA molecule makes any information concerning its properties
potentially important.  Unfortunately, a number of intriguing
experimental observations in this field could not be accurately
interpreted in terms of DNA structure because the direct NMR and X-ray
methods are not always applicable and often limited in accuracy.
Molecular dynamics simulations of the type described here can help to
clarify some long standing issues in this domain provided that the
mechanism of A\lrar B transitions is similar {\em in silico} and {\em
in vitro}.

Under low hydration, dynamic DNA structure systematically converges to
the A-form conformation that is very similar for different sequences
and very close to the canonical conformation observed in fiber
crystals.  In contrast, a much more significant and sequence dependent
divergence from experimental structures is systematically obtained for
B-DNA.  There is a surprising correspondence between this difference
and experimental data. The high regularity and the absence of sequence
effect for A-form was noticed long ago for X-ray fiber diffraction
patterns \cite{Leslie:83} and later confirmed in the ensemble of single
crystal A-DNA structures \cite{Suzuki:97}. Thus, in both experiment and
simulations, the A-form of DNA is virtually insensitive to the base
pair sequence. In simulations it appears also to be relatively
insensitive to the accuracy of the forcefield that do not allow MD
trajectories to come closer to experimental B-DNA conformations.
These observations may have a common physical origin.  Notably, they
can be attributed to the A-form being dominated by the "electrostatic
sandwich" in the major groove.  The strong interactions of phosphate
groups with metal cations effectively impose geometric constraints
upon the inter-phosphate distances and suppress all other factors that
might affect the overall structure.  Similar types of interactions are
also possible for the A-form stretches observed in complexes with
proteins.  In all such structures resolved by now the major groove of
A-DNA is exposed to solvent whereas extensive protein DNA contacts are
observed in the minor groove.

The systematic deviations of the computed B-DNA conformations from the
canonical one are similar to the recent simulations for analogous
sequences \cite{Trantirek:00,Stefl:01,Lankas:02}. The Cornell et al parameters
\cite{Cornell:95} have a well documented tendency to underestimate the
average helical twist \cite{Cheatham:99}, and for the sequences
studied here this bias is perhaps the largest. For poly(dG).poly(dC)
this results in structures with very A-like helical parameters.  At
the same time, sugar pseudorotation dynamics samples mainly from
South/East phases characteristic of B-DNA (see \rfg{g12sgph}). Earlier
such structures were interpreted as strongly underwound B-DNA or as
A-DNA with B-like backbone \cite{Trantirek:00,Stefl:01,Lankas:02}.
Unfortunately, it is rather difficult to tell exactly how strong is
the deviation of these conformations from experimental data. They are
rather different from G-tracts found in a large number of A-DNA single
crystal structures. However, these structures are found in a unique
A-DNA packing characterized by very special DNA-DNA interactions in
wide minor grooves \cite{Wahl:99}. The same trend is not seen in short
G-tracts found in protein-DNA complexes. For instance, one of the
resolved structures of the Y-family DNA polymerase contains a DNA
fragment with an exposed GGGGG terminus \cite{Ling:01}.  This G-tract
features a normal B-DNA conformation with South sugar puckers and the
average Twist of 34.2\degree rather close to that of generic B-DNA.
Unfortunately, the last value is much larger than the measured
solution value (ca 32.7\degree\ for long G-tracts \cite{Peck:81})
suggesting that the DNA structure may again be affected by the crystal
environment.  On the other hand, solution NMR and Circular dichroism
(CD) studies indicate that the G-tract duplexes computed with the
Cornell et al force field resemble experimental structures
\cite{Trantirek:00,Stefl:01}. The corresponding CD spectra are very
similar to typical A-DNA suggesting that the high hydration
conformation of poly(dG).poly(dC) already has an A-like base pair
stacking. At the same time, the sugar phases determined by NMR are all
at the South. These conformations can hardly be assigned to the A-form
because a cooperative transition to a genuine A-form is still
distinguishable in solution as well as in crystalline fibers although
the corresponding spectral changes are small
\cite{Trantirek:00,Stefl:01,Arnott:74}. These subtle changes can well
result from a transition pattern similar to that in \rfg{g12alpa},
with sugar phases switching to the North and grooves changing their
width, but relatively minor shifts in helical parameters.

The poly(dA).poly(dT) conformation at the high hydration limit are
much closer to the canonical B-DNA structure although, in agreement
with earlier reports \cite{McConnell:01}, it is also underwound by ca
4\degree\ with respect to the experimental solution value
\cite{Peck:81,Rhodes:81,Strauss:81}. Our simulations managed to
reproduce the well-documented relative A-form phobicity of this
sequence in several different aspects. Notably, there is a well
defined range of hydration numbers where poly(dA).poly(dT) and
poly(dG).poly(dC) are stable in B- and A-forms, respectively.
Dodecamer A12  resembles  poly(dA).poly(dT) properties in crystalline
fibers in that it refuses to go to the A-form and is prone to
denaturation. At the same time, longer poly(dA).poly(dT) fragments
with GC termini adopt the A-form without great problems, which agrees
with experiments in solution \cite{Tolstorukov:01}.  It seems rather
perplexing that a B\rar A transition for A12 could not be obtained
in spite of the fact that its A-form was found to be perfectly stable
under low hydration. It is known, however, that in some experimental
conditions, the B\rar A transition in DNA can go via temporary
condensation or denaturation \cite{Vorlickova:91b}. If the lowest
energy transition pathway goes like that in our conditions, the
molecular dynamics simulations would not be able to reach the final
state because of the limited duration of trajectories.  One should
also keep in mind that the experimental B\lrar A transitions in DNA
reportedly involve macroscopically slow relaxation phases and
hysteresis effects \cite{Falk:62,Lindsay:88}.  Even though they
usually concern long DNA chains, they may be due to local effects like
slow ion diffusion especially under low hydration.

Our results feature the complexity of the poly(dA).poly(dT) DNA as
regards the B/A polymorphism, but they leave unanswered the question
concerning the main physical origin of such extraordinary behavior. A
number of factors can contribute to the poly(dA).poly(dT) persistence
with respect to the B\rar A transition.  The sequence dependent
propensities towards A and B-forms should certainly originate from of
the specific base stacking \cite{Calladine:84,Mazur:89,Hunter:93}.
However, the variety and the character of the observed environmental
effects suggest that subtle structural deviations prompted by specific
base-base contacts are amplified by strong interactions with water and
counterions. For instance, the narrow minor groove profile in the
B-form of poly(dA).poly(dT) may be sealed by the hydration spine
structure \cite{Dickerson:82}.  The same feature should produce a
twofold effect upon the counterion distribution around DNA
\cite{Rouzina:98,Hud:03}.  The free positive ions should tend to
accumulate in front of the narrow minor groove, and, simultaneously,
their concentration in the major groove is reduced. Both these
features should stabilize the B-form.  Earlier experimental and
computational observations also suggest that a special role should be
played by thymine methyl groups. They hinder a negative slide movement
involved in the B\rar A transition \cite{Hunter:93} and form a
continuous non-polar cluster in the major groove of B-DNA that should
provide additional hydrophobic stabilization \cite{Tolstorukov:01}.
The same feature effectively reduces the accessible volume of the
major groove and can prevent accumulation of free solvent cations
\cite{Mzjacs:03}.

Many years ago, the contrasting propensities of poly(dA).poly(dT) and
poly(dG).poly(dC) DNA to adopt A- and B-forms presented the first
experimental demonstration of sequence dependent properties of the
double helical DNA structure \cite{Pilet:75}. Since then the repertory
of reported sequence effects has many times increased, and yet the
exact physical origin of this particular difference remains
controversial. It is shown here that A\lrar B transitions observed in
water drop simulations exhibit clear trends qualitatively similar to
the long known experimental observations.  These results corroborate
the putative general role of the intra-duplex electrostatic
condensation mechanism for A\lrar B transitions in DNA {\em in vitro}
and suggest that future studies in the same direction can give more
definite answers to the issues discussed here.

\section*{Methods}

A series of MD simulations was carried out for double helical
dodecamer fragments poly(dA).poly(dT) and poly(dG).poly(dC) referred
to as A12 and G12, respectively, as well as a few derivatives of these
sequences. Water drops of 300, 400, 500, 600, 700, 800, 1000, and 2000
molecules were used, which gives approximate hydration numbers
(water/nucleotide) 13, 17, 21, 25, 29, 33, 42, and 83, respectively.
Additional simulations with a hydration number around 171 were carried
out by using periodical boundary conditions. Below for brevity the
simulation conditions outlined above are referred to as HN13, HN17,
and so fourth. All calculations were continued long enough to obtain
convergent average parameters of the final states. In this way a
quasi-static pattern of B\lrar A transitions can be reproduced
resembling {\em in vitro} titration experiments

The simulation protocols were similar to the earlier water drop
simulations \cite{Mzjacs:02,Mzjacs:03}. We use the internal coordinate
molecular dynamics (ICMD) method \cite{Mzjcc:97,Mzbook:01} adapted for
DNA \cite{Mzjacs:98,Mzjchp:99} with the time step of 0.01 ps. In this
approach, the DNA molecule has all bond length and almost all bond
angles fixed at their standard values. The only variable bond angles
are those centered at the sugar C1',...,C4', and O4' atoms, which
assures the flexibility of the furanose rings. In contrast, bases,
thymine methyls, and phosphate groups move as articulated rigid
bodies, with only rotations around single bonds allowed. The highest
frequencies in thus obtained models are additionally balanced by
increasing rotational inertia of the lightest rigid bodies as
described earlier\cite{Mzjpc:98,Mzjacs:98}.  The possible physical
effects of the above modifications have been discussed elsewhere
\cite{Mzbook:01,Mzctpc:01}. The electrostatic interactions are treated
with the recent version of the SPME method \cite{Essmann:95}
specifically adapted for infinite vacuum boundary conditions
\cite{Mzjacs:02}. The common values of Ewald parameters were used,
that is 9 \AA\ truncation for the real space sum and $\beta\approx 0.35$.
Reference simulations with periodical boundaries were carried out as
described before \cite{Mzjacs:02}, with the standard SPME method in
NVT ensemble conditions with a rectangular unit cell of
$45\times45\times65$ \AA\ under normal water density.

The initial states were prepared as follows. The canonical A and
B-DNA \cite{Arnott:72} were used as standard initial conformations.
The DNA molecule was first immersed in a large rectangular TIP3P
\cite{Jorgensen:83} water box of and next external solvent molecules
were removed by using a spherical distance cut-off from DNA atoms. The
cut-off radius was adjusted to obtain the desired number of water
molecules remaining. The drop was neutralized by randomly placing
Na$^+$ ions at water positions selected so that their distances from
DNA were 5 \AA\ or larger. The initial counterion distribution was
pre-equilibrated by running 1 ns dynamics in water drops of 500
molecules for A-DNA and 800 molecules for B-DNA, with DNA atoms weakly
restrained to their initial positions. The final drop size was
adjusted by adding or removing water from the surface. The described
procedure was intended to assure the start of dynamics from closely
similar states regardless of the drop size.

Every system was energy minimized first with the solute held rigid and
then with all degrees of freedom. Dynamics were initiated with the
Maxwell distribution of generalized momenta at low temperature. The
system was next slowly heated to 250 K and equilibrated during several
picoseconds.  Production trajectories were computed with the
temperature bound to 293 K by the Berendsen algorithm
\cite{Berendsen:84} with a relaxation time of 10 ps. For better
comparison with earlier simulations of A\lrar B transitions, the
original Cornell et al. force filed \cite{Cornell:95} was used.
Duration of production runs varied from 2 to 25 ns depending upon the
observed character of dynamics. The conformations were saved with a
2.5 ps interval. The results were analyzed with in-house routines and
the Curves program \cite{Curves:}.

\bibliography{preprint}
\bibliographystyle{jmb}

\end{document}